\documentclass[useAMS,usenatbib]{mn2e}

\usepackage{times,graphicx,amssymb,natbib}

\title[The Jeans Mass and Disc Fragmentation]{The Jeans Mass as a Fundamental Measure of Self-Gravitating Disc Fragmentation and Initial Fragment Mass}
\author[Duncan Forgan and Ken Rice]{Duncan Forgan $^{1}$\thanks{E-mail:
dhf@roe.ac.uk} and Ken Rice$^{1}$ \\
$^{1}$Scottish Universities Physics Alliance (SUPA), Institute for Astronomy, University of Edinburgh, Blackford Hill, Edinburgh, EH9 3HJ, Scotland, UK}
\begin{document}

\date{Accepted}

\pagerange{\pageref{firstpage}--\pageref{lastpage}} \pubyear{}

\maketitle

\label{firstpage}

\begin{abstract}

\noindent As a formation route for objects such as giant planets and low-mass stars in protostellar discs (as well as stars in AGN discs), theories of self-gravitating disc fragmentation need to be able to predict the initial masses of fragments.  We describe a means by which the local Jeans mass inside the spiral structure of a self-gravitating disc can be estimated.  If such a self-gravitating disc satisfies the criteria for disc fragmentation, this estimate provides a lower limit for the initial mass of any fragments formed. We apply this approach to a series of self-gravitating protostellar disc models, to map out the typical masses of fragments produced by this formation mode.  We find a minimum fragment mass of around 3 Jupiter masses, which is insensitive to the stellar mass, and that - within the parameter space surveyed - fragments with masses between 10 and 20 Jupiter masses are the most common.  We also describe how the Jeans mass allows us to derive a more general criterion for disc fragmentation, which accounts for the processes of viscous heating, radiative cooling, accretion and the disc's thermal history.  We demonstrate how such a criterion can be determined, and show that in limiting cases it recovers several fragmentation criteria that have been posited in the past, including the minimum cooling time/maximum stress criterion.

\end{abstract}


\section{Introduction}

\noindent During the process of low mass star formation, an excess of angular momentum in the parent molecular cloud will generally result in the formation of a circumstellar disc.  In the early phases of this disc's existence, it is expected to be relatively massive, cool, and extremely weakly ionised.  Therefore, disc self-gravity is expected to play a dominant role in protostellar and protoplanetary disc evolution \citep{Lin1987,Laughlin1994}, in the same manner it is expected to play a role in the evolution of Active Galactic Nuclei (AGN), which also possess accretion discs \citep{Shlosman1989,Goodman2003}.  A self-gravitating disc becomes gravitationally unstable if \citep{Toomre_1964}:

\begin{equation} Q = \frac{c_s \kappa}{\pi G \Sigma} \sim 1, \end{equation}

\noindent where $c_s$ is the local sound speed, $\Sigma$ is the surface density and $\kappa$ is the epicyclic frequency (for Keplerian discs, this is equal to the angular frequency $\Omega$).  This result applies primarily to axisymmetric perturbations, and is usually extended to non-axisymmetric perturbations by changing the critical $Q$ value to $Q< 1.5-1.7$, a result established empirically by numerical simulations (see \citealt{Durisen_review} for a review).  

The onset of gravitational instability (GI) produces non-axisymmetric, spiral structures.  It has been shown that self-gravitating discs will typically settle into what is known as a marginally stable state, where the stress heating produced by the spiral waves is balanced (at least approximately) by radiative cooling \citep{Paczynski1978}.  In the marginally stable state, the spiral waves become quasi-steady structures that, while individually transient, as a whole remain for long timescales, mediating the transport of angular momentum outward (and the transport of mass inward).

The evolution of self-gravitating discs can therefore be neatly parametrised using the phenomenology of viscosity, much in the same way that the action of magneto-rotational instability (MRI) has been parametrised in less massive, more ionised discs \citep{Blaes1994,Balbus1999}.  This pseudo-viscous parametrisation, and its encapsulation in the dimensionless $\alpha$ parameter \citep{Shakura_Sunyaev_73}, has been a useful tool in simplifying the computation of the structure of self-gravitating discs \citep{Rafikov_05,Clarke_09,Rice_and_Armitage_09}, as well as computing the evolution of said discs \citep{Lynden-Bell1974,Pringle1981a,Lin1990,Rice2010}.  

While useful, this pseudo-viscous approximation is precisely that - an approximation, which is accurate only for a limited region of the available disc parameter space.  In particular, for the approximation to be valid, the angular momentum transport that occurs must be local.  Gravity is an inherently non-local force, and if the spiral structures in the disc can exert sufficient gravitational field stresses on fluid elements at large separations, then the pseudo-viscous approximation begins to fail, and the disc's angular momentum transport is governed by global properties \citep{Balbus1999}.  The breakdown of the local approximation has been shown to occur when the disc-to-star mass ratio, $q$, is greater than 0.5, and/or the disc scale height satisfies $H/r > 0.1$ \citep{Lodato_and_Rice_04,Forgan2011}.  Discs which satisfy these conditions typically possess strong $m=2$ spiral structures, which exert significant torques from large distances, and exhibit highly variable temperature structures, and consequently highly variable $Q$.  We will see that the issue of whether disc angular momentum transport is local or global will come to be of great importance in this paper.

If the transport is indeed local, the balance between the magnitude of disc stresses and the magnitude of radiative cooling has also been useful in describing the fragmentation of local self-gravitating discs.  \citet{Gammie} demonstrated that a second criterion, alongside the $Q$-criterion above, must be satisfied if a self-gravitating disc is to fragment.  This criterion is related to the local cooling time, $t_{cool}$, and is usually expressed using the dimensionless cooling time parameter $\beta_{c}=t_{cool} \Omega$, i.e.

\begin{equation} \beta_{c} \leq \beta_{crit}, \end{equation}

\noindent where the exact value of $\beta_{crit}$ is typically established by numerical simulations.  Using the 2D shearing sheet approximation, \citet{Gammie} showed that when the ratio of specific heats in two dimensions $\gamma_{2D}=2$, $\beta_{crit} = 3$.  \citet{Rice_et_al_05} extended this analysis to 3D smoothed particle hydrodynamics (SPH) simulations (where $\beta_{c}$ was fixed at some value), and explored the dependence of $\beta_{crit}$ on $\gamma$.  They were able to confirm Gammie's assertion that, in the case where the disc has reached thermal equilibrium (i.e. the radiative cooling and stress heating due to gravitational instability are in balance), the fragmentation criterion can be recast in terms of the $\alpha$ parameter:

\begin{equation} \alpha = \frac{4}{9 \gamma(\gamma-1)\beta_c}, \end{equation} 

\noindent where the critical value of $\alpha=\alpha_{crit}\approx 0.06$ for all values of $\gamma$.  The fragmentation criterion can now be interpreted as a maximum GI stress that the disc can support in thermal equilibrium without fragmentation.  

The inner regions of discs are typically too hot for $Q\sim 1$, and are therefore not gravitationally unstable and incapable of fragmentation.  Also, if the primary contribution to $\alpha$ is from GI stress, it will typically decrease with increasing proximity to the central star \citep{Armitage_et_al_01,Rice_and_Armitage_09,Rice2010}.  These facts immediately suggest a minimum radius at which fragmentation can occur.  Many authors, using a variety of techniques, including semi-analytic methods, grid-based and particle based simulations, find that for self-gravitating protostellar discs the minimum is approximately $40$ au, using typical disc parameters \citep{Rafikov_05, Matzner_Levin_05, Whit_Stam_06,Mejia_3,Stamatellos2008, intro_hybrid, Clarke_09}.  

This also constrains the maximum available mass to form a fragment (as typically equilibrium surface density profiles decrease with increasing distance).  Despite such constraints, the consensus view is that the objects formed occupy the higher mass end of the planetary regime (and the low mass end of the stellar regime).  \citet{Kratter2009} estimate the available mass within the most unstable wavelength to fragmentation (the disc scale height $H$) using one-dimensional disc models to obtain fragment masses, and then model the fragments' growth embedded in the disc to constrain expected planet masses.  Studies of the behaviour of $\beta_c$ (or more strictly $t_{cool}$) in similar disc models \citep{Rafikov_05, Nero2009} tend to agree with \citet{Kratter2009} in that they calculate a minimum fragment mass of around $5 M_{\rm Jup}$, where $M_{\rm Jup}$ is the mass of Jupiter.

While the $Q$ and $\beta_c$ criteria provide an elegant means by which to predict disc fragmentation, there are complicating factors that must also be considered.  Perhaps the most important is the fact that (for the most part) these numerically simulated discs are \emph{isolated}.  We expect that self-gravitating discs - as they form at an early phase in protostellar evolution - will be surrounded by a substantial envelope from which the disc can accrete.  The potential for mass-loading as a result of accretion can alter the local optical depth as well as the local stresses felt by the disc.  \citet{Kratter2010a} show in grid-based numerical simulations that systems accreting rapidly can sustain $\alpha \sim 1$ without fragmenting.  This is also confirmed by \citet{Harsono2011} using SPH simulations, where they ascribe the extra Reynolds stresses to velocity shear at the disc surface.  Both show that the disc infall rate and disc-to-star accretion rate are typically proportional to each other, and that non-local angular momentum transport is significant.  These factors alone show the limited validity of isolated, constant $\beta_c$ simulations in estimating disc fragmentation criteria.

Further to this, the imposition of a fixed $\beta_c$ is somewhat unphysical.  As the disc cools through varying opacity regimes, the opacity is expected to change significantly (see e.g. \citealt{intro_hybrid, Cossins2008}).  \citet{Clarke2007} carry out SPH simulations where $\beta_c$ is initially large, and gradually decreased towards the critical value.  They find that the eventual $\beta_{crit}$ can be up to a factor of 2 lower.  This reduction of $\beta_{crit}$ will increase $\alpha_{crit}$ by the same factor - while the results of \citet{Rice_et_al_05} provide a useful watermark, the exact value of $\alpha_{crit}$ is not precisely determined.  Indeed, a related result of \citet{Cossins2010} shows that $\beta_{crit}$ also depends on the opacity regime the gas is in, producing isolated discs which require $\alpha_{crit}=0.1$ for fragmentation.

Taking these factors into consideration, it becomes clear that the current use of $\beta_c$ or $\alpha$ is not sufficiently general as a fragmentation criterion.  While it is suitable for self-gravitating discs of constant $\beta_c$ that do not accrete, such discs are not typically found in astrophysical situations.  Ideally, we would like to move towards a situation where the fragmentation criterion is sufficiently rich in its content that it can describe the full complexity of the fragmentation process in realistic self-gravitating discs.

This paper introduces two useful analytical formalisms: the first describes the local Jeans mass in a marginally stable, self-gravitating disc (which is described in section \ref{sec:jeansmass}); this provides a means of predicting the masses of objects formed as a result of disc fragmentation.  In section \ref{sec:betaj} we demonstrate that this expression for the Jeans mass can be used to develop a new timescale criterion, which folds in the influences of stress heating, radiative cooling, accretion and thermal history in general.  Having developed these concepts, we construct simple one-dimensional models to investigate the typical fragment masses produced by self-gravitating discs (described in section \ref{sec:method}), the results and discussion of which can be seen in section \ref{sec:results}.  We summarise the work in section \ref{sec:conclusions}.

\section{The Jeans Mass inside a spiral wave perturbation} \label{sec:jeansmass}

\noindent While it is perhaps more appropriate to look at the Toomre mass in this rotating configuration, the Jeans and Toomre masses are related by factors of order unity when the disc is marginally stable \citep{Nelson_res}, so it is safe to use the more conceptually simple Jeans mass.  As fragments will typically form thanks to the perturbing influence of a spiral density wave, it is sensible to look at the Jeans mass inside the wave itself.  

Let us assume that we have a marginally stable disc ($Q \sim 1$), with some local surface density $\Sigma$, producing spiral wave perturbations with fractional amplitude $\frac{\Delta \Sigma}{\Sigma}$.  Essentially, we want a perturbation to contain a mass $M$ greater than the local Jeans Mass, $M_J$.  If we assume a spherical Jeans Mass:

\begin{equation} M_J = \frac{4}{3} \pi \left(\frac{\pi c^2_s}{G\rho_{pert}}\right)^{3/2} \rho_{pert} = \frac{4}{3} \pi^{5/2} \frac{c^3_s}{G^{3/2} \rho^{1/2}_{pert}}, \end{equation}

\noindent where $\rho_{pert}$ is the density of the perturbation, and $c_s$ is the local sound speed.  In the midplane, 

\begin{equation}\rho_{pert} = \Sigma_{pert}/2H,  \end{equation}

\noindent where by definition

\begin{equation} \Sigma_{pert} = \Sigma\left(1+ \frac{\Delta \Sigma}{\Sigma}\right). \end{equation}  

\noindent Substituting into $M_J$ gives

\begin{equation} M_J = \frac{4\sqrt{2}}{3} \pi^{5/2} \frac{c^3_s  H^{1/2} }{G^{3/2} \Sigma^{1/2}\left(1 + \frac{\Delta \Sigma}{\Sigma}\right)}. \end{equation}

\noindent We now eliminate $\Sigma$ using $Q$: 

\begin{equation} \left(G\Sigma\right)^{1/2} = \left(\frac{c_s \Omega}{\pi Q}\right)^{1/2}. \end{equation}  

\noindent This gives

\begin{equation} M_J = \frac{4\sqrt{2}\pi^3 Q^{1/2}}{3G}\frac{c^{5/2}_s H^{1/2} }{\Omega^{1/2}\left(1 + \frac{\Delta \Sigma}{\Sigma}\right)}. \end{equation}

\noindent Using $H = c_s/\Omega$, we can eliminate $\Omega$ to obtain: 

\begin{equation} M_J = \frac{4\sqrt{2}\pi^3 }{3G}\frac{Q^{1/2} c^2_s H}{\left(1 + \frac{\Delta \Sigma}{\Sigma}\right)}. \label{eq:mjeans_nobeta}\end{equation}

\noindent This is a general description of the Jeans Mass inside the spiral wave, achieved using the following assumptions only:

\begin{enumerate}
 \item The disc is marginally stable ($Q \sim 1$)
 \item $\rho = \Sigma / 2H$ at the midplane
 \item $H = c_s/\Omega$
 \item The Jeans mass is spherical
\end{enumerate}

\noindent Using the non self-gravitating prescription for $H$ might seem hazardous, however for marginally stable discs

\begin{equation} H_{nsg} = \frac{c_s}{\Omega} \approx H_{sg} = \frac{c^2_s}{\pi G \Sigma}, \end{equation}

\noindent (this can be confirmed by substituting for $Q \sim 1$).  Assumption 2 is reasonable for cases where the thin disc approximation applies (as is the case for assumption 3).  We can calculate the fractional amplitude $\Delta \Sigma/\Sigma$ using the empirical results of \citet{Cossins2008}:

\begin{equation} \frac{\Delta \Sigma}{\Sigma} = \frac{1}{\sqrt{\beta_c}}. \end{equation}

\noindent This now gives

\begin{equation} M_J = \frac{4\sqrt{2}\pi^3 }{3G}\frac{Q^{1/2} c^2_s H}{\left(1 + 1/\sqrt{\beta_c}\right)}. \label{eq:mjeans_H} \end{equation}

\noindent This quantity is now completely calculable using azimuthally averaged variables.  The Jeans mass depends on the local sound speed, the angular velocity and the thermal physics of the gas (which is sensitive to the surface density and temperature structure).  We must however note that using this empirical prescription for the fractional amplitude requires that the angular momentum transport in the disc be local.  

\section{A General Timescale Criterion for Disc Fragmentation} \label{sec:betaj}

\noindent Can the Jeans mass provide a timescale criterion for disc fragmentation (much as the cooling time has done in the past)?  Let us now define three timescales, normalised to the orbital period:

\begin{equation} \Gamma_J = \frac{M_J}{\dot{M}_J} \Omega ,\end{equation}
\begin{equation} \beta_c = t_{cool} \Omega  ,\end{equation}
\begin{equation} \Gamma_\Sigma = \frac{\Sigma}{\dot{\Sigma}} \Omega .\end{equation}

\noindent We use $\beta$ and $\Gamma$ to distinguish between variables of differing behaviour.  $\beta_c$ is positive definite - this is to keep it in line with the conventional definition as it is currently used (and also to prevent $\Delta \Sigma/\Sigma$ taking imaginary values).  $\Gamma_\Sigma$ measures the competition between disc accretion and stellar accretion. $\Gamma_J$ measures the timescale on which the local Jeans Mass changes - in $\Gamma_J$ orbital periods, we can expect the Jeans mass to either double or decrease to zero, depending on the sign of $\Gamma_J$.  $\dot{\Sigma}$ and $\dot{M}_J$ can be either positive or negative - for fragmentation to be favourable, the local Jeans mass must be decreasing.  A small, negative $\Gamma_J$ therefore represents the most likely circumstances for disc fragmentation.  The critical value of $\Gamma_J$ is less clear - any disc that can maintain a negative $\Gamma_J$ will proceed towards fragmentation.  Rather than presenting us with two discrete regimes, fragmenting and non-fragmenting, we see a spectrum of potential outcomes, some fragmenting rapidly; some fragmenting on much longer timescales (possibly longer than the lifetime of the disc, and therefore effectively non-fragmenting); and others moving away from potential fragmentation (either slowly or rapidly).  We suggest that $-10< \Gamma_J < 0$ gives highly favourable conditions for prompt fragmentation, but the lower limit is by no means fixed, and will require empirical confirmation. 

We wish to now derive $\Gamma_J$: substituting for the self-gravitating scale height, we can rewrite the Jeans mass in terms of $\Sigma$, $c_s$ and $\beta_c$:

\begin{equation} M_J = \frac{4\sqrt{2}\pi^2 }{3G^2}\frac{Q^{1/2} c^4_s}{\Sigma\left(1 + 1/\sqrt{\beta_c}\right)}. \label{eq:mjeans_sigma} \end{equation}

\noindent To derive $\dot{M}_J$, we can use the chain rule:

\begin{equation} \dot{M}_J = \frac{\partial M_J}{\partial c_s}\dot{c}_s + \frac{\partial M_J}{\partial \Sigma}\dot{\Sigma} + \frac{\partial M_J}{\partial \beta_c}\dot{\beta_c}. \label{eq:MJdot} \end{equation}

\noindent Instead of directly calculating $\dot{c}_s$, it is easier to describe the rate of change of specific internal energy $u$, which we can relate back to $c_s$:

\begin{equation} c^2_s = \gamma(\gamma-1)u. \end{equation}

\noindent We will assume that the sound speed varies due to two effects only - firstly, radiative cooling, given by

\begin{equation} \dot{u}_{cool} = - \frac{u}{t_{cool}} \end{equation}

\noindent and secondly, heating due to viscous dissipation, given by 

\begin{equation} \dot{u}_{heat} = 9/4 \alpha \gamma (\gamma-1) u \Omega \end{equation}

\noindent Using the chain rule on $\dot{c}_s$ and rearranging gives:

\begin{equation} \dot{c}_s = \frac{d c_s}{du} \dot{u}= 1/2 c_s\left(9/4 \alpha \gamma(\gamma-1)\Omega - \frac{1}{t_{cool}}\right). \end{equation}

\noindent We now use equation (\ref{eq:mjeans_sigma}) to evaluate equation (\ref{eq:MJdot}), where we subsequently substitute back for $M_J$:

\begin{displaymath} \dot{M}_J = \frac{4}{c_s} M_J \frac{c_s}{2}\left(\frac{9\alpha \gamma(\gamma-1)\Omega}{4} - \frac{1}{t_{cool}}\right) - \frac{M_J}{\Sigma} \dot{\Sigma} \end{displaymath}
\begin{equation} + \frac{M_J}{4\beta^{3/2}_c\left(1+1/\sqrt{\beta_c}\right)}\dot{\beta_c}. \end{equation}

\noindent Equivalently,

\begin{displaymath} \dot{M}_J = M_J \Omega \left(2\left(-\frac{1}{\beta_c} + \frac{9\alpha \gamma(\gamma-1) }{4}\right) - \frac{1}{\Gamma_\Sigma} \right. \end{displaymath}
\begin{equation} \left. + \frac{\dot{\beta_c}}{4\beta^{3/2}_c\Omega\left(1+1/\sqrt{\beta_c}\right)}\right). \end{equation}

\noindent This quickly gives the Jeans timescale $\Gamma_J$ as

\begin{displaymath} \Gamma_J = \left(2\left(-\frac{1}{\beta_c} + \frac{9\alpha \gamma(\gamma-1)}{4}\right) - \frac{1}{\Gamma_\Sigma} \right.\end{displaymath}
\begin{equation} \left. + \frac{\dot{\beta_c}}{4\beta^{3/2}_c\Omega\left(1+1/\sqrt{\beta_c}\right)}\right)^{-1}. \label{eq:betaj}\end{equation}

\noindent Remember that we require the magnitude of $\Gamma_J$ to be small - making it more negative means that the Jeans mass will decrease on a slower timescale, making fragmentation less likely.  Also, small positive values of $\Gamma_J$ will rapidly increase the Jeans mass, suppressing fragmentation.  \\

\noindent To recap, the following assumptions have been made:

\begin{enumerate}
 \item The disc is marginally stable ($Q \sim 1$)
 \item $\rho = \Sigma / 2H$ at the midplane
 \item $H = c_s/\Omega = c^2_s/\pi G \Sigma$
 \item $ \Delta \Sigma/\Sigma = 1/\sqrt{\beta_c}$ (which implicitly assumes locally transporting discs)
 \item The Jeans mass is spherical 
\end{enumerate}

\subsection{The Limiting Values of $\Gamma_J$ - Recovering Older Criteria}

\subsubsection{The ``Standard'' Cooling Time Criterion}

\noindent If we assume $\dot{\Sigma}=0$, then $\frac{1}{\Gamma_\Sigma} \rightarrow 0$.  If we also assume $\dot{\beta_c} = 0$, we recover

\begin{equation} \Gamma_J = \left(2\left(-\frac{1}{\beta_c} + 9/4 \alpha \gamma(\gamma-1)\right) \right)^{-1}, \end{equation}

\noindent and the fragmentation is governed by the cooling rate (or equivalently the disc's GI stresses).  The minus sign appears due to our definition of $\beta_c$: this is sensible considering that cooling will reduce the local Jeans Mass.  If we substitute for $\alpha$ in the case of thermal equilibrium

\begin{equation} \alpha = \frac{4}{9 \gamma(\gamma-1)\beta_c}, \end{equation} 

\noindent then the cooling and heating are in detailed balance and $\Gamma_J \rightarrow \infty$, i.e. $M_J = const.$.  This is essentially a re-statement of the assumed thermodynamic equilibrium, and is almost axiomatic.  If the gravitational torque saturates at some maximum possible value of $\alpha$, then continually increasing the cooling rate will eventually force $\Gamma_J$ to become small and negative, rapidly decreasing the local Jeans mass until fragmentation becomes possible even for relatively low density perturbations.  We must once more note that local, pseudo-viscous transport is assumed.  If the disc angular momentum transport is global, we must replace the heating term with a more complicated expression reflecting non-local effects.  Indeed, even discs with local angular momentum transport will be subject to non-local heating such as from stellar irradiation - this extra heating term should also be considered if modelling/simulation permits.

\subsubsection{The Effect of Thermal History}

\noindent If we allow $\dot{\beta_c}$ to be non-zero, then we can compare two limiting cases. If the change in cooling time is low, then we are in the regime described in the previous section.  If however the change in cooling time is rapid:

\begin{equation} \dot{\beta_c} >> \beta_c \Omega: \,\,\,\rightarrow \,\,\, \Gamma_J \approx \frac{4\Omega \beta^{3/2}_c\left(1+1/\sqrt{\beta_c}\right)}{\dot{\beta_c}} ,\end{equation}

\noindent with $\Gamma_J$ becoming small.  This is in accordance with \citet{Clarke2007}, who show that discs with low $\dot{\beta_c}$ are generally more stable to fragmentation than those with high (negative) $\dot{\beta_c}$.  But what is responsible for the disc's fragmentation? We can investigate this by applying \citet{Clarke2007}'s prescription for $\beta_c(t)$:

\begin{equation} \beta_c(t) = \beta_c(0)\left(1- \frac{t}{T}\right) , \,\, t<T\end{equation}

\noindent where $T$ defines the magnitude of $\dot{\beta_c}$:

\begin{equation} \dot{\beta_c} = -\frac{\beta_c(0)}{T} \end{equation}

\noindent Substituting into equation (\ref{eq:betaj}), this gives

\begin{displaymath} \Gamma_J = \left( -\frac{1}{\beta_c(0) (1 - t/T)} + \frac{9\alpha \gamma (\gamma-1)}{4} -  \right. \end{displaymath}
\begin{equation} \left. \frac{1}{4 T\Omega(\sqrt{ \beta_c(0)(1-t/T)} + 1) }\right)^{-1} \end{equation}

\noindent Let us consider two cases: if there is no maximum stress beyond which fragmentation occurs, then the first two terms will always be able to balance, $\beta_c$ will eventually tend towards zero by definition, and

\begin{equation} \Gamma_J \rightarrow - 4T\Omega \end{equation}

\noindent If $\dot{\beta_c}$ is large, $T$ will be small and hence $\Gamma_J$ will be small and negative, promoting fragmentation.  If there \emph{is} a maximum stress which the disc exceeds, thermal equilibrium is no longer possible, and $\Gamma_J$ will continue to decrease towards small values.  If we fix the viscous heating term at its maximum, we can recover the rate of change of $\Gamma_J$ against time in this scenario:

\begin{equation} \frac{d \Gamma_J}{dt} = \frac{\beta^2_J\beta_c(0)}{T\Omega} \left(\frac{1}{\beta^2_c} + \frac{1}{8T\Omega\sqrt{\beta_c}\left(\sqrt{\beta_c}+1\right)^2}\right) \label{eq:dbetajdy}\end{equation}

\noindent The derivative is positive, and its magnitude depends primarily on the rate of change of $\beta_c$ and on $\Gamma_J$ itself.  For discs that are rapidly changing their cooling time, $T$ is small, and hence $\Gamma_J$ will increase rapidly towards fragmentation in only a few orbits.  For discs changing their cooling time on longer scales, then $T$ and $\Gamma_J$ are both initially large, and therefore the derivative is also initially large.  However, as the derivative decreases rapidly with decreasing $\Gamma_J$, large $T$ discs will proceed much more slowly towards fragmentation, requiring many more orbits.  In fact, the number of orbits required for fragmentation in this case will depend linearly on $T$ - increasing $T$ by a factor of 100 will require 100 times as many orbits to witness fragmentation.  If $T$ is sufficiently large, then secular evolution of the disc (e.g. significant stellar accretion) can act to reduce the likelihood of fragmentation by removing mass from the system.

A similar result applies to simulations where $\beta_c$ is reduced and then held at some value $\beta_{hold}$ (such as \citealt{Clarke2007}).  This action will fix $\Gamma_J$ at some value given by the mismatch between the first and second terms in equation (\ref{eq:betaj}).  Low values of $\beta_{hold}$ will increase the mismatch, decreasing $\Gamma_J$ and reducing the timescale for fragmentation to occur.  Given this and equation (\ref{eq:dbetajdy}), we can now make some general statements about the work of \citet{Clarke2007}.  They see some discs fragmenting at very low $\beta_c$ because once favourable fragmentation conditions arise, there is a finite timescale for fragmentation to occur, defined by $\Gamma_J$.  If $T$ is sufficiently small (to the point of being unphysical), then $\beta_c$ can become extremely small before $\Gamma_J$ reaches values necessary for fragmentation. Also, we show that as \citet{Clarke2007} note themselves, if they were able to run their high $T$ simulations with $\beta_c$ below the critical value for a sufficiently long time, they would also probably fragment.  The length of simulation time required to produce a fragment will be calculable from the evolution of $\Gamma_J$.   While high $T$ simulations may be more stable to fragmentation in the short-term, their long-term stability is not likely to be much better than simulations where the cooling time is rapidly changing.

As a final aside on the subject, we must be careful about the sign of $\dot{\beta_c}$.  Fragmentation is more likely if $\Gamma_J$ is negative, and hence $\dot{\beta}_c$ should be negative also.  If the change in cooling time is rapid and \emph{positive}, then $\beta_c$ will become too large, and the Jeans mass will either increase or become steady at a higher value (which can be seen from the $\beta_c$ dependence of equation (\ref{eq:mjeans_sigma})).  Large fluctuations in $\beta_c$ are therefore not a guarantee of fragmentation - \emph{we require the cooling rate to increase rapidly for fragmentation to be more favourable}.  This is akin to the fragmentation of shocked flows producing star formation in molecular clouds (e.g. \citealt{Heitsch2008,Bonnell2008}), and may explain the formation of objects during protostellar encounters with sufficiently extended discs \citep{Thies2005,Shen2010,Thies2010}.  Equally, this accounts for fragmentation suppression during encounters with compact discs \citep{Lodato_encounters,encounters}, where the subsequent increase in optical depth ensures $\dot{\beta_c}$ is large and positive.

\subsubsection{Accretion as a Shortcut to Fragmentation}

\noindent Let us assume for the moment that heating and cooling are in exact balance (with a large, constant value of $\beta_c$), such that the heating and cooling terms in equation \ref{eq:betaj} cancel and $\dot{\beta}_c=0$, and hence we can approximate

\begin{equation} \Gamma_J = -\Gamma_\Sigma, \end{equation}

\noindent i.e. glibly speaking, discs which have a higher accretion rate are more favourable to fragmentation.  Of course, accretion of envelope material does not merely increase the disc mass - it affects the local thermodynamics and the angular momentum distribution, both of which can push the disc away from $Q\sim 1$.  We should therefore also demand 

\begin{equation} \frac{dQ}{d\Sigma} = \frac{\partial Q}{\partial c_s}\frac{d c_s}{d \Sigma} + \frac{\partial Q}{\partial \Omega}\frac{d \Omega}{d \Sigma} + \frac{\partial Q}{\partial \Sigma} \lesssim 0 \end{equation}

\noindent if the disc is to remain amenable to fragmentation.  Substituting $d c_s / d\Sigma = \dot{c}_s/ \dot{\Sigma}$ (and similarly for $\Omega$), we obtain

\begin{equation} \frac{dQ}{d\Sigma} = \frac{Q}{c_s}\frac{\dot{c}_s}{\dot{\Sigma}} +  \frac{Q}{\Omega}\frac{\dot{\Omega}}{\dot{\Sigma}} -\frac{Q}{\Sigma} \lesssim 0\end{equation}

\noindent Upon rearrangement, the following condition now appears:

\begin{equation} \frac{Q}{\dot{\Sigma}} \left(\frac{\dot{c}_s}{c_s} + \frac{\dot{\Omega}}{\Omega} - \frac{\dot{\Sigma}}{\Sigma} \right) \lesssim 0 \end{equation}

\noindent At the critical point (where $\frac{dQ}{d\Sigma}=0$), this can be simplified to merely

\begin{equation}  \frac{\dot{\Sigma}}{\Sigma} \gtrsim \frac{\dot{c}_s}{c_s} + \frac{\dot{\Omega}}{\Omega}. \end{equation}

\noindent If accretion results in greater deposition of angular momentum or in local heating, then the disc moves away from fragmentation (this is intuitively obvious from the definition of $Q$).  Also, it further underlines the stabilising influence of sustained heating, either by accretion shocks or by stellar irradiation \citep{Rafikov_07,Mejia_4,Stamatellos2008}.

\subsubsection{Summary}

\noindent Marginally stable self-gravitating discs are susceptible to fragmentation if:

\begin{enumerate}
 \item The cooling rate is sufficiently high (or equivalently the disc gravitational stresses are sufficiently high), or
 \item The cooling rate increases sufficiently rapidly (or the disc gravitational stresses increase rapidly), or
 \item The disc's envelope accretion rate is sufficiently high (while maintaining $Q\sim 1$).
\end{enumerate}

\subsubsection{Predicting Fragment Masses}

\noindent We should be able to identify an equilibrium Jeans Mass by demanding $\dot{M}_J=0 $, or

\begin{equation} -\frac{1}{\beta_c} + 9/4 \alpha \gamma(\gamma-1) - \frac{1}{\Gamma_\Sigma} + \frac{\dot{\beta_c}}{4\beta^{3/2}_c\Omega\left(1+1/\sqrt{\beta_c}\right)} = 0 \end{equation}

\noindent This is a first order non-linear ODE for $\beta_c$ (assuming a constant $\Gamma_\Sigma$).  In the constant $\beta_c$ formalisms of the past, this becomes

\begin{equation} \beta_c = 9/4 \alpha \gamma(\gamma-1) - \Gamma_\Sigma \end{equation}

\noindent These formalisms are not predictors of the fragment mass \emph{per se}, but they show the secular Jeans mass, and to some extent the secular fragment mass if the disc is left to its own devices and satisfies the correct criteria.  Also, in practice the cooling rate will be a function of deposition rate (in the sense that this maintains a local optical depth), the angular velocity will be a function of mass ratio and the cooling rate will be sensitive to changes in temperature and surface density.  This differential equation is only soluble numerically, and probably only with numerical simulations.

In the case of simple analytic models, it is more expedient to impose the fragmentation criteria, and calculate $M_J$ at the fragmentation boundary.  This method will now be applied to such disc models in the following sections.

\section{Fragment Masses in Simple Disc Models with Local Angular Momentum Transport}\label{sec:method}

\begin{figure*}
\begin{center}$
\begin{array}{c}
\includegraphics[scale = 0.5]{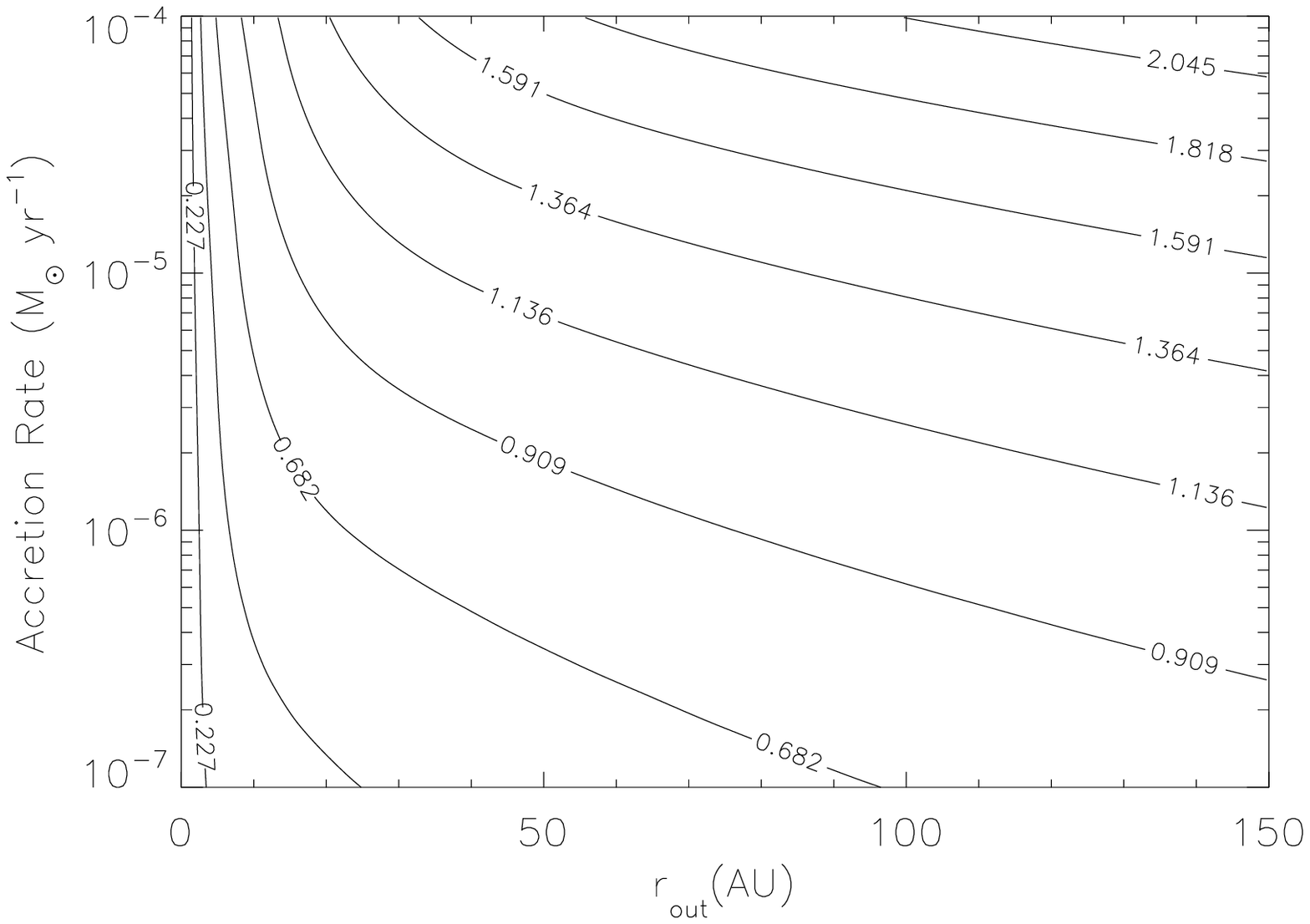} \\
\includegraphics[scale = 0.5]{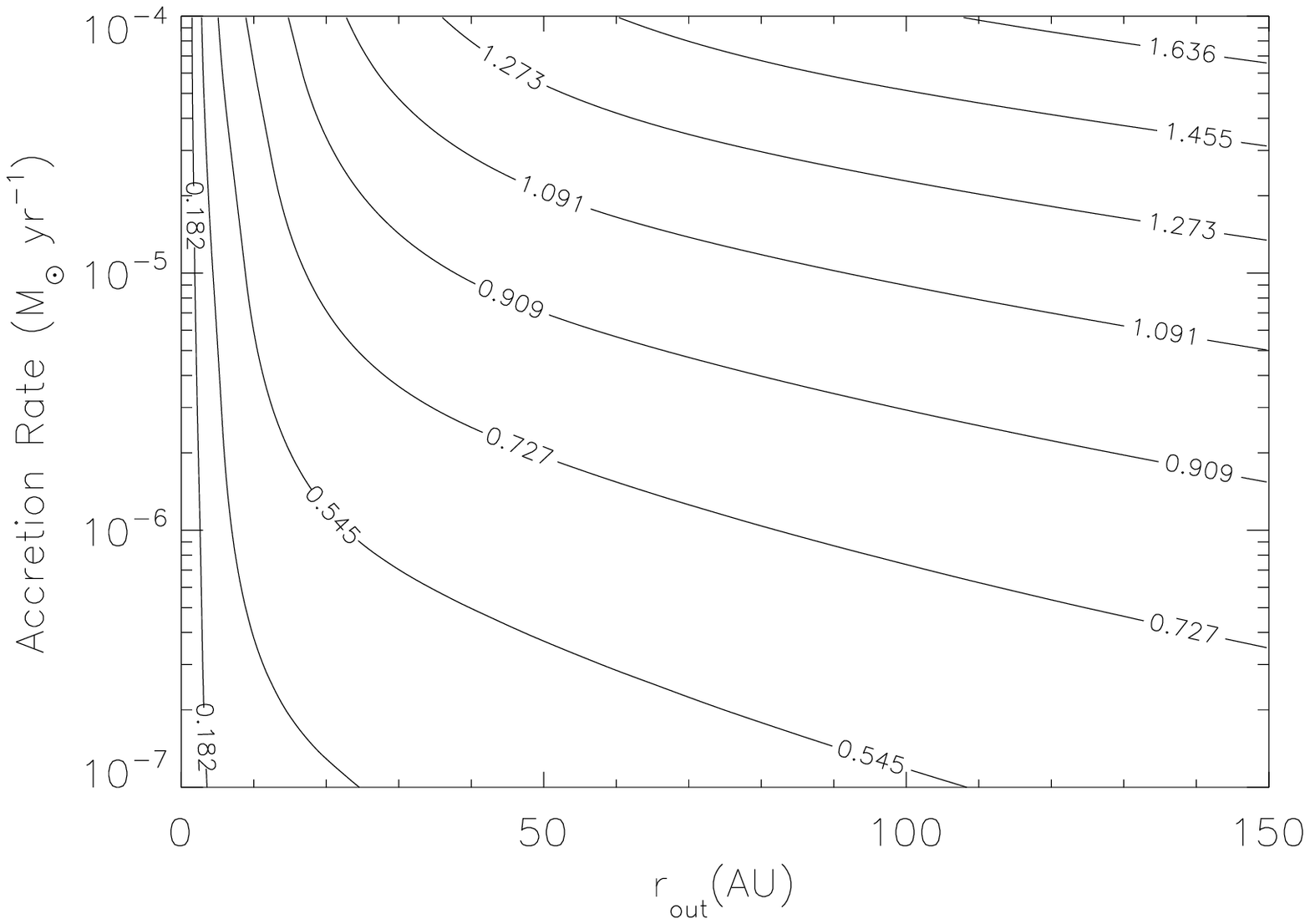} \\
\includegraphics[scale = 0.5]{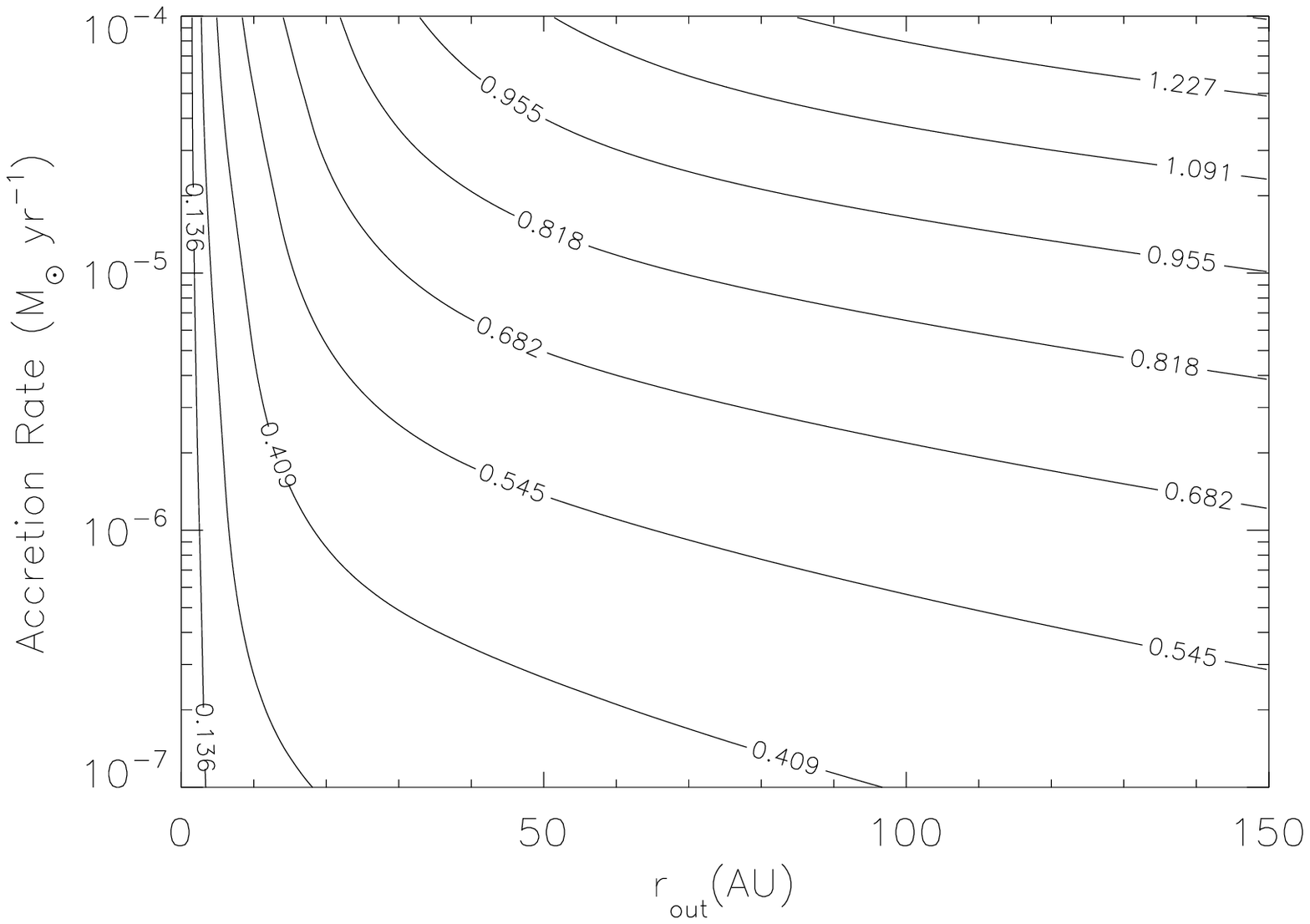} \\
\end{array}$
\caption{2D contour plots of the disc to star mass ratio, as a function of the steady state `pseudo-viscous' accretion rate $\dot{M}$, and the disc's outer radius $r_{\rm out}$, for stellar masses of  $M_{*} = 0.5 M_{\odot}$ (top), $M_{*} = 1 M_{\odot}$ (middle) and  $M_{*} = 2 M_{\odot}$ (bottom).}\label{fig:mdot_r_q}
\end{center}
\end{figure*}

\noindent To investigate the dependence of the Jeans mass on disc parameters, we construct a model of a self-gravitating disc assuming a steady state accretion rate $\dot{M}$ and an outer disc radius $r_{\rm out}$.  We take an approach essentially identical to that of \citet{Levin2007}, \citet{Clarke_09} and \citet{Rice_and_Armitage_09}.  For the sake of simplicity, we assume $\dot{\Sigma}=0$, $\dot{\beta_c}$ is small, and impose $-5<\Gamma_J <0$ as the  criterion for fragmentation.  Cautioned by the results of \citet{Cossins2010}, we impose a saturating value for $\alpha = \alpha_{sat}=0.1$.

The model is constructed as follows: we assume the disc is Keplerian and gravitationally unstable at all radii, i.e.

\begin{equation} Q = \frac{c_s \Omega}{\pi G \Sigma} \sim 1 \label{eq:Q}\end{equation}

\noindent for any value of $r$.  The steady state accretion rate is 

\begin{equation} \dot{M} = \frac{3\pi\alpha c^2_s \Sigma}{\Omega}, \label{eq:mdot}\end{equation}

\noindent which is assumed to be constant over all radii.  We will also assume that $\alpha$ is determined by the constraint of local thermal equilibrium, i.e.

\begin{equation} \alpha = \frac{4}{9\gamma(\gamma-1)\beta_c}, \label{eq:alpha}\end{equation}

\noindent where $\beta_c = \left(u/\dot{u}\right) \Omega$ is a function of volume density $\rho$ and temperature.  This assumes of course that the disc is local -  we will leave a discussion of the validity of this $\alpha$ parametrisation for a later section.  The cooling function is 

\begin{equation} \dot{u} = \frac{\sigma_{SB} T^4}{\tau + 1/\tau}, \end{equation}

\noindent where $\sigma_{SB}$ is the Stefan-Boltzmann constant, and the optical depth $\tau = \Sigma \kappa(\rho,c_s)$, with $\kappa$ the opacity.   Our aim is to calculate the three unknowns $\alpha$, $c_s$ and $\Sigma$ at any radius.  Equations (\ref{eq:Q}), (\ref{eq:mdot}) and (\ref{eq:alpha}) place three constraints, and the system is therefore soluble.  We employ the equation of state of \citet{Stam_2007} (see also \citealt{intro_hybrid,Cossins2008,Rice_and_Armitage_09}) to determine the opacity, the ratio of specific heats ($\gamma$) and the mean molecular weight ($\mu$) as a function of volume density and sound speed.  

This allows us to rapidly produce a suite of disc models spanning a broad parameter space in $\dot{M}$ and $r_{\rm out}$.  Figure \ref{fig:mdot_r_q} shows the required disc-to-star mass ratios for disc models with a given $\dot{M}$ and $r_{\rm out}$.  It is important to note that quite high values of $q$ are required to sustain self-gravitating discs - for typical T Tauri values (say $\dot{M} = 10^{-8} M_{\odot}\,yr^{-1}$, $r_{\rm out} = 100 $ au), the disc-to-star mass ratio required is $q\sim0.25$, several orders of magnitude higher than observations suggest \citep{Andrews2007,Greaves2010}.  If the local temperature exceeds $1000$ K, then the ionisation fraction should be sufficiently high for MRI to operate, and enhance the local accretion rate.  We model this by setting $\alpha = \alpha_{MRI} = 0.01$, and re-calculate $\Sigma$ such that the steady-state accretion rate is obtained.  While this differs from the expected episodic nature of MRI activation in this scenario \citep{Armitage_et_al_01,Zhu2009}, its draining effect on the resulting disc mass is important.  For modest outer radii, the disc-to-star mass ratio increases to over 0.5 for all stellar masses.  Even with MRI activation providing drainage, the discs remain sufficiently cool that large amounts of matter can still be found in the inner regions (this is due to $\alpha$ decreasing steeply towards the central object).

Models which satisfy the $\Gamma_J$ criterion are classified as fragmenting models, and the local Jeans Mass is then calculated using equation (\ref{eq:mjeans_H}).  We investigate three different stellar masses, generating 250,000 models for each.  The models span accretion rates from $\dot{M} = 10^{-7} - 10^{-4} M_{\odot}\,yr^{-1}$, and disc outer radii from $1.3-150$ au (the inner disc radius is assumed to be $1$ au for all models).

\section{Results \& Discussion} \label{sec:results}

\begin{figure*}
\begin{center}$
\begin{array}{c}
\includegraphics[scale = 0.5]{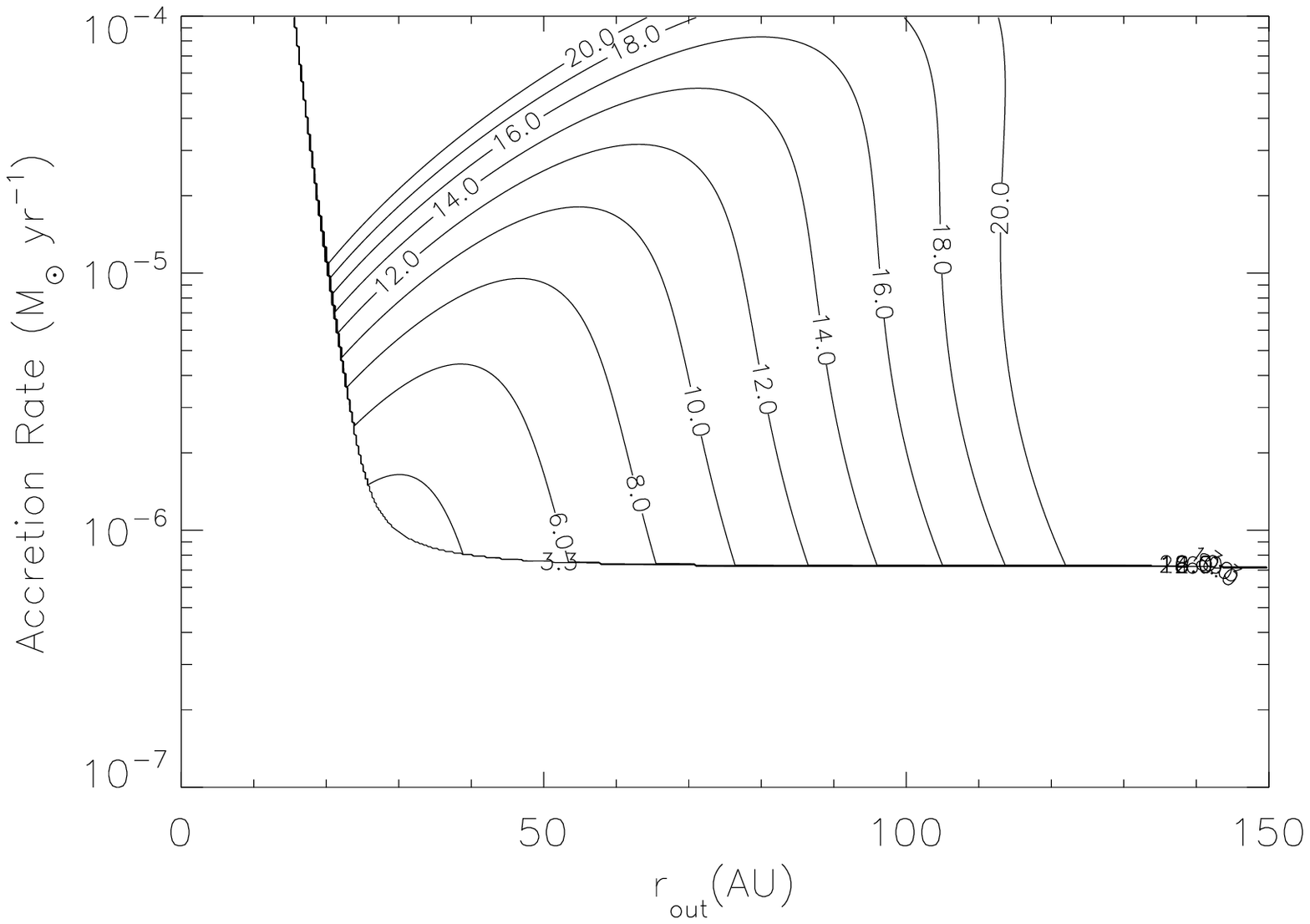} \\
\includegraphics[scale = 0.5]{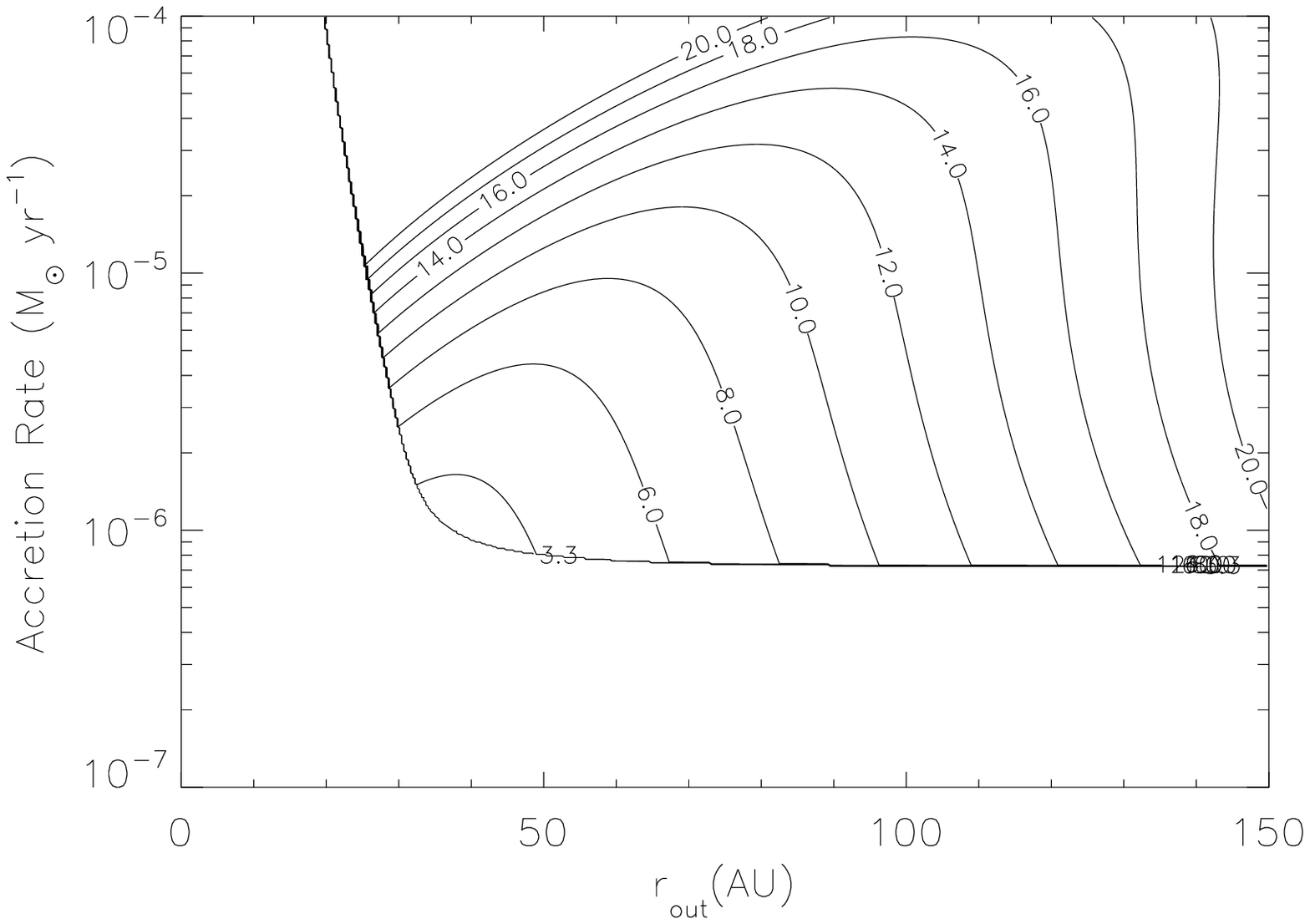} \\
\includegraphics[scale = 0.5]{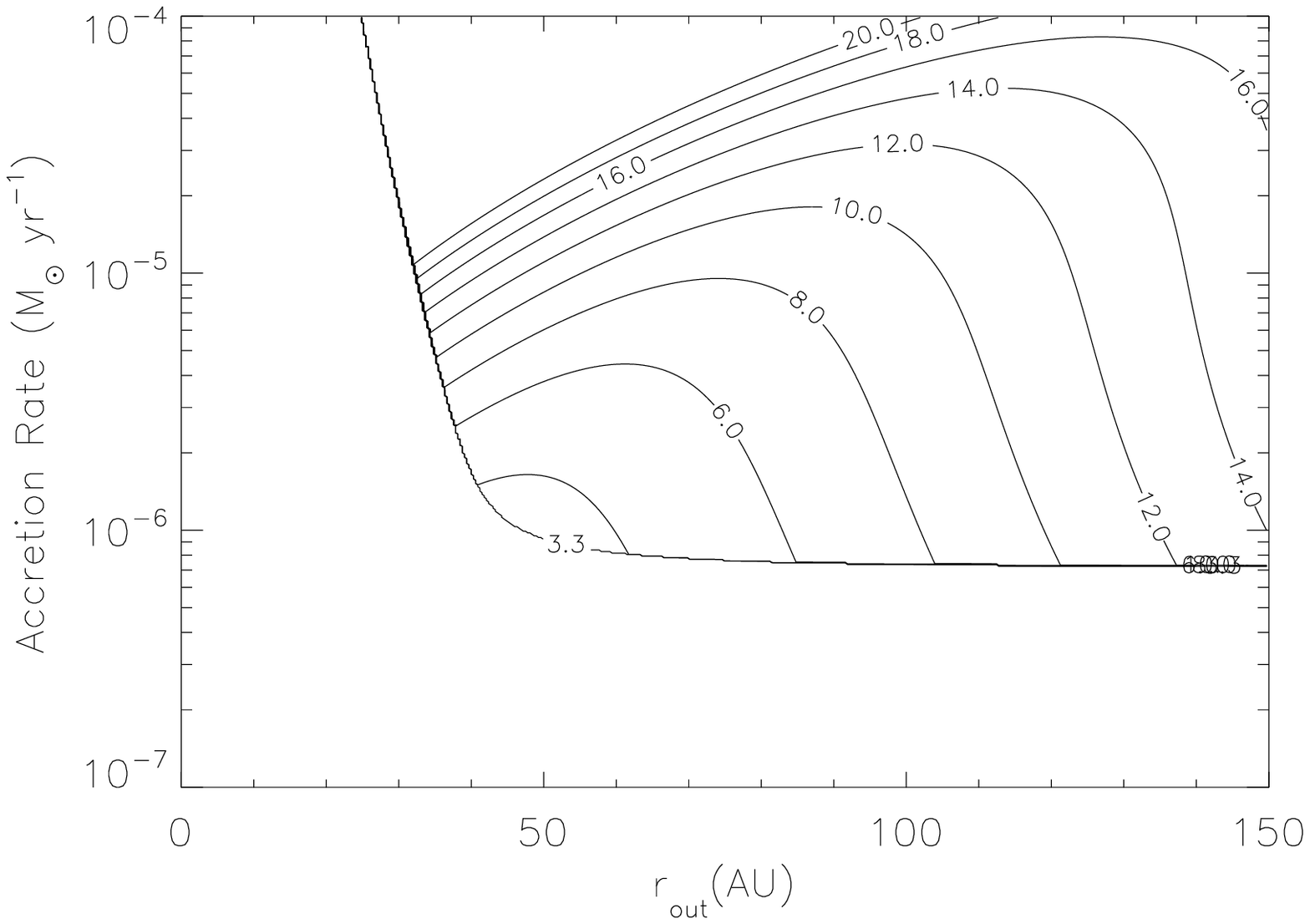} \\
\end{array}$
\caption{2D contour plots of the expected Jeans mass in spiral perturbations (in Jupiter masses), as a function of the steady state `pseudo-viscous' accretion rate $\dot{M}$, and the disc's outer radius $r_{\rm out}$, for stellar masses of  $M_{*} = 0.5 M_{\odot}$ (top), $M_{*} = 1 M_{\odot}$ (middle) and  $M_{*} = 2 M_{\odot}$ (bottom).}\label{fig:mdot_r_frag}
\end{center}
\end{figure*}

\noindent Figure \ref{fig:mdot_r_frag} shows the expected Jeans Mass at fragmentation as a function of $\dot{M}$ and the disc outer radius $r_{\rm out}$.  We present data for three different stellar masses: $0.5$, $1$ and $2 M_{\odot}$.  The minimum Jeans mass across all stellar masses remains roughly constant at around $3 \, M_{\rm Jup}$, as does the maximum Jeans Mass ($\sim 125 \, M_{\rm Jup}$, or $0.1 M_{\odot}$).  These discs are defined to be quasi-steady with constant disc mass and local temperature - the minimum Jeans mass is therefore fixed by the heating/cooling balance in the disc (and by extension the local opacity).

The minimum mass agrees well with \citet{Kratter2009} (where they state a similar value for the minimum fragment mass at a temperature of 10 K and an accretion rate of $10^{-7}  M_{\odot} yr^{-1}$), as well as the results of \citet{Rafikov_05} and \citet{Nero2009}.  This demonstrates once more that objects formed by disc fragmentation will not occupy the Earth-mass regime (at least initially).  If disc fragmentation is to produce such objects, the subsequent evolution of the fragments must include significant mass loss, as well as strong core sedimentation.  While it has been shown that fragments of mass $< 1 M_{\rm Jup}$ are slow to produce substantial cores by sedimentation \citep{Helled2008}, the typical fragments produced in these models may be more successful, especially with the dust-concentrating effects of the spiral structure induced by gravitational instability \citep{Rice2004,Clarke2009,Boley2010a,Nayakshin2010a}.

As has been found many times previously, discs with outer radii less than around 40 au will typically not form fragments \citep{Rafikov_05, Matzner_Levin_05, Whit_Stam_06,Mejia_3,Stamatellos2008, intro_hybrid, Clarke_09}.  While this is more a symptom of the imposed disc model (which in itself is not a particularly new implementation), we can use this result to satisfy ourselves that the model is indeed correctly distinguishing the fragmentation boundary.  The dependence of the radial fragmentation boundary on stellar mass (i.e. lower mass stars have a lower boundary) is also an indication that the model is indeed performing as expected.

The distribution of the contours shows a clear trend as the stellar mass is increased - the $M_* = 0.5 M_{\odot}$ star has contours more closely packed along the $r$-axis, and is able to form (for example) a $20\,M_{\rm Jup}$ fragment at $120$ au (at the lowest $\dot{M}$ which permits fragmentation at that radius).  By contrast, the $M_* = 2 M_{\odot}$ star can only form a $10\,M_{\rm Jup}$ fragment at the same location in $\dot{M}-r$ space.  This is sensible, given that the disc around the $M_* = 2 M_{\odot}$ star has a much lower mass ratio for those model parameters.  

\begin{figure*}
\begin{center}$
\begin{array}{c}
\includegraphics[scale = 0.5]{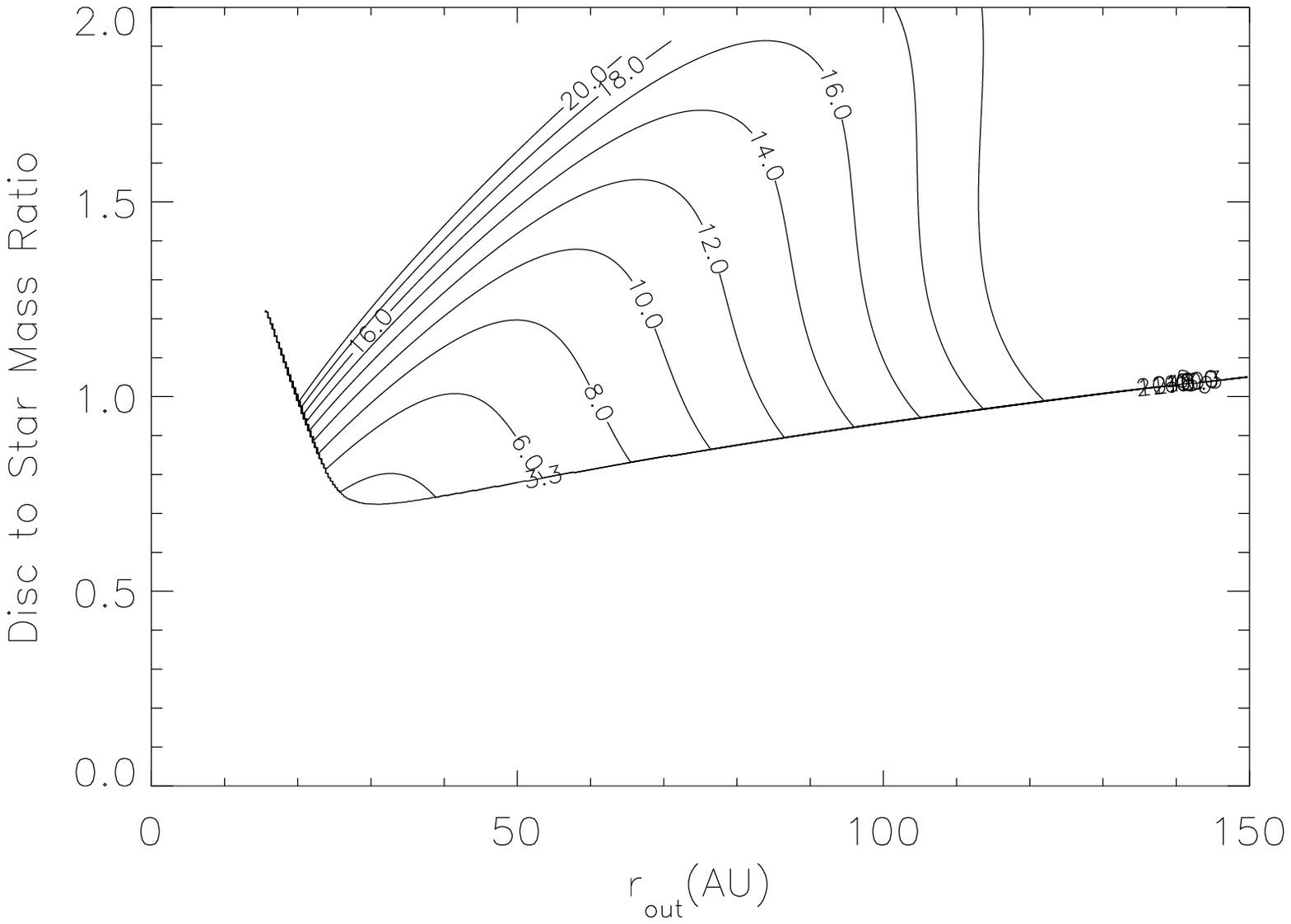} \\
\includegraphics[scale = 0.5]{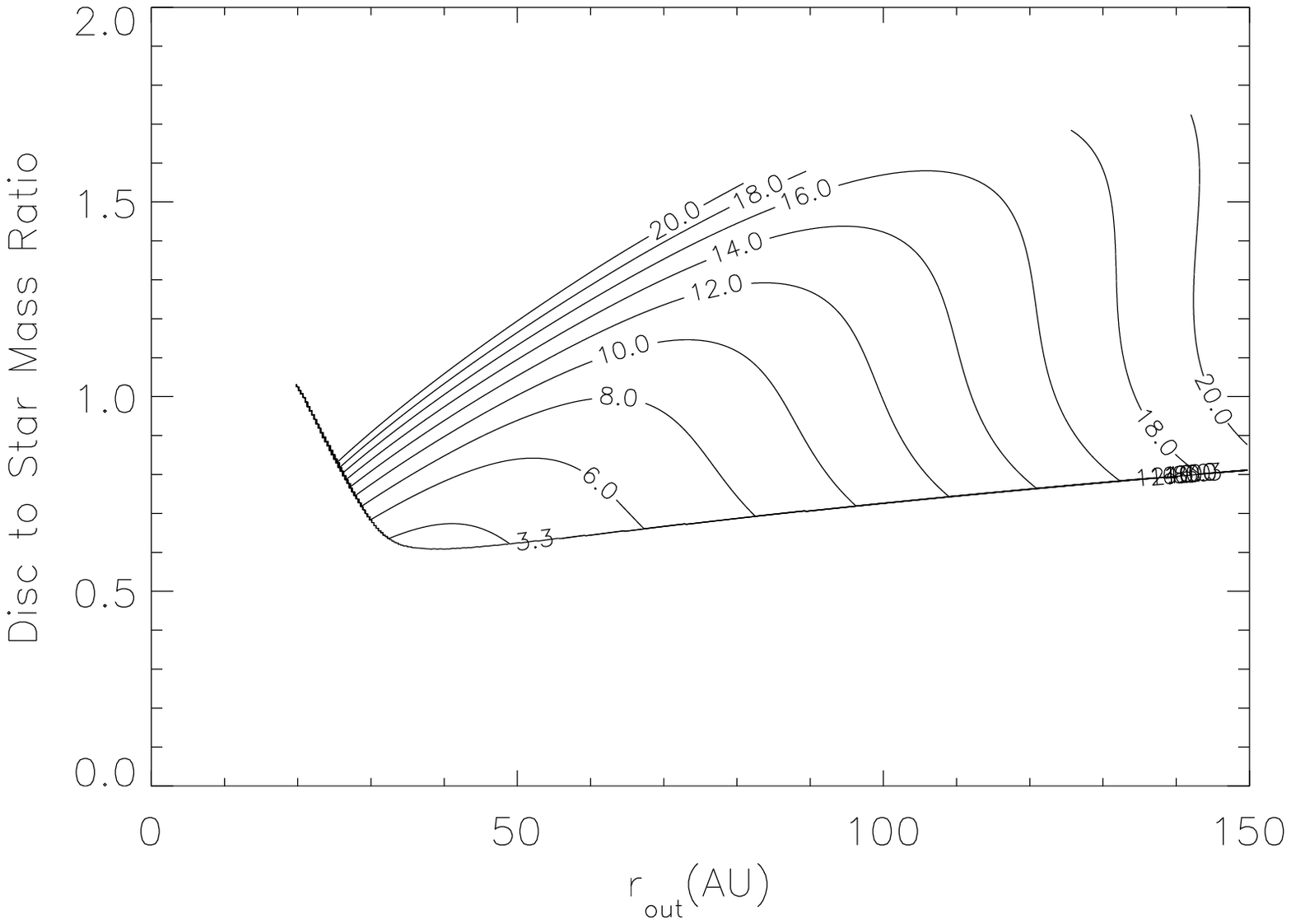} \\
\includegraphics[scale = 0.5]{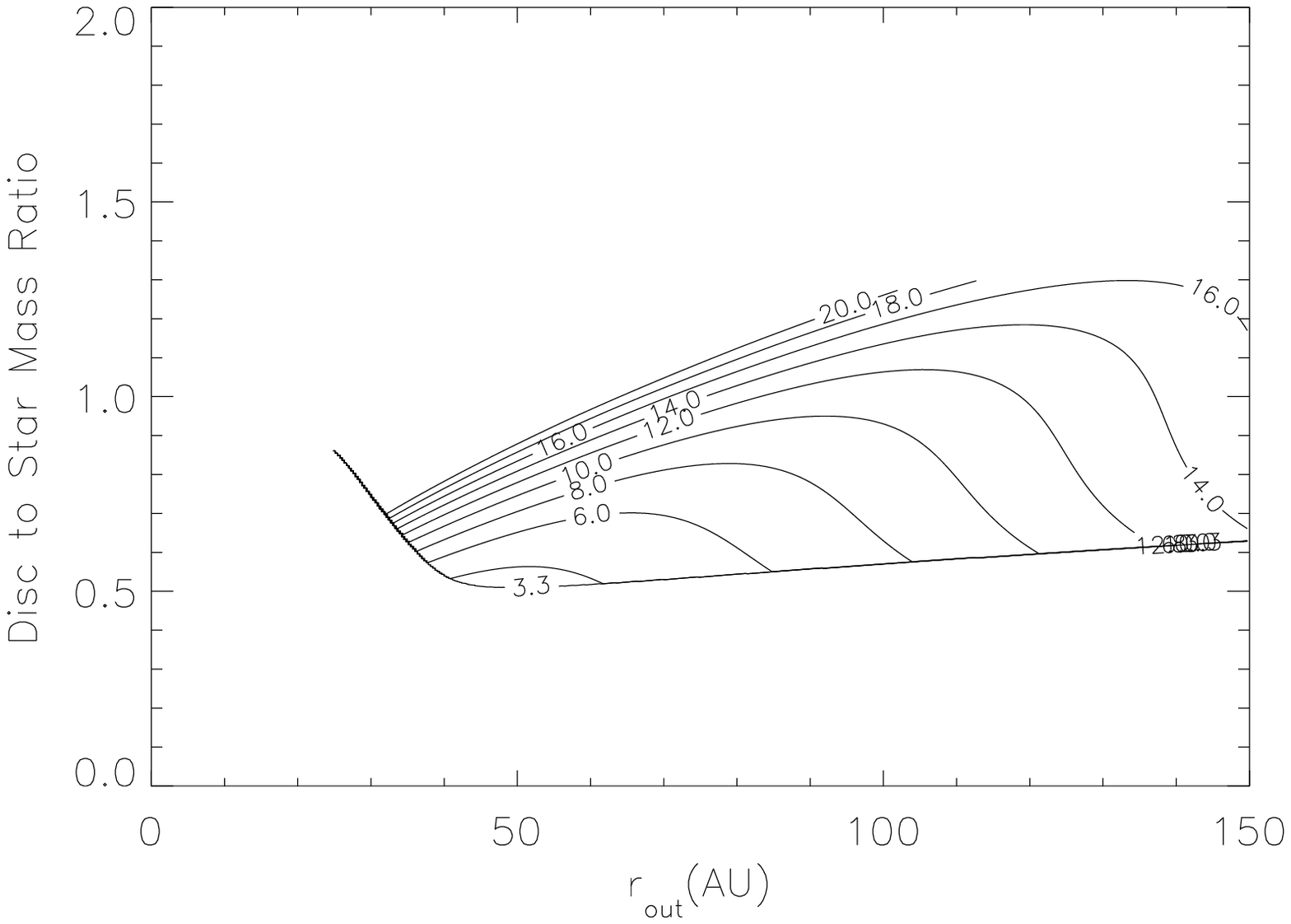} \\
\end{array}$
\caption{2D contour plots of the expected Jeans mass in spiral perturbations (in Jupiter masses), as a function of the disc-to-star mass ratio $q$, and the disc's outer radius $r_{\rm out}$, for stellar masses of  $M_{*} = 0.5 M_{\odot}$ (top), $M_{*} = 1 M_{\odot}$ (middle) and  $M_{*} = 2 M_{\odot}$ (bottom).}\label{fig:q_r_frag}
\end{center}
\end{figure*}

This can be seen in Figure \ref{fig:q_r_frag}, which shows the expected Jeans mass (in Jupiter Masses) at fragmentation, as a function of the disc-to-star mass ratio, $q$.  We calculate models for which $q$ is very much higher than observed.  Indeed, current modelling indicates that $q\geq 1$ discs can only exist at very early times \citep{Machida2010}, and they evolve rapidly towards states where $q<0.5$ \citep{Forgan2011}.  The fact that typically $q>0.5$ is required for fragmentation strains our assumption of local transport to breaking point.  Further to this, requiring such large values of $q$ further boosts the argument for rapid giant planet formation while the system is still young \citep{Greaves2010}.

If we consider discs with $q=1$ and $r_{\rm out} = 50$ au, then it becomes clear that the $M_*= 2 M_{\odot}$ star can form the most massive fragments ($> 20 M_{\rm Jup}$), whereas the $M_* = 0.5 M_{\odot}$ star can only form fragments of mass $\approx 7 M_{\rm Jup}$, with the $M_*= 1M_{\odot}$ star forming fragments of around $10 M_{\rm Jup}$.

In the same fashion, the discs around the $M_* = 0.5 M_{\odot}$ star require $q\approx 1.25$ to form a $10 \, M_{\rm Jup}$ fragment, while discs around the $M_* = 2 M_{\odot}$ require $q\approx 0.75$ (for $r_{\rm out} = 50$ au).  We can identify a general trend - discs around higher mass stars can satisfy a maximum $\alpha$ fragmentation criterion (and hence form a minimum fragment mass of $\sim 3 M_{\rm Jup}$) at lower $q$ than discs around lower mass stars.

\begin{figure*}
\begin{center}$
\begin{array}{cc}
\includegraphics[scale = 0.5]{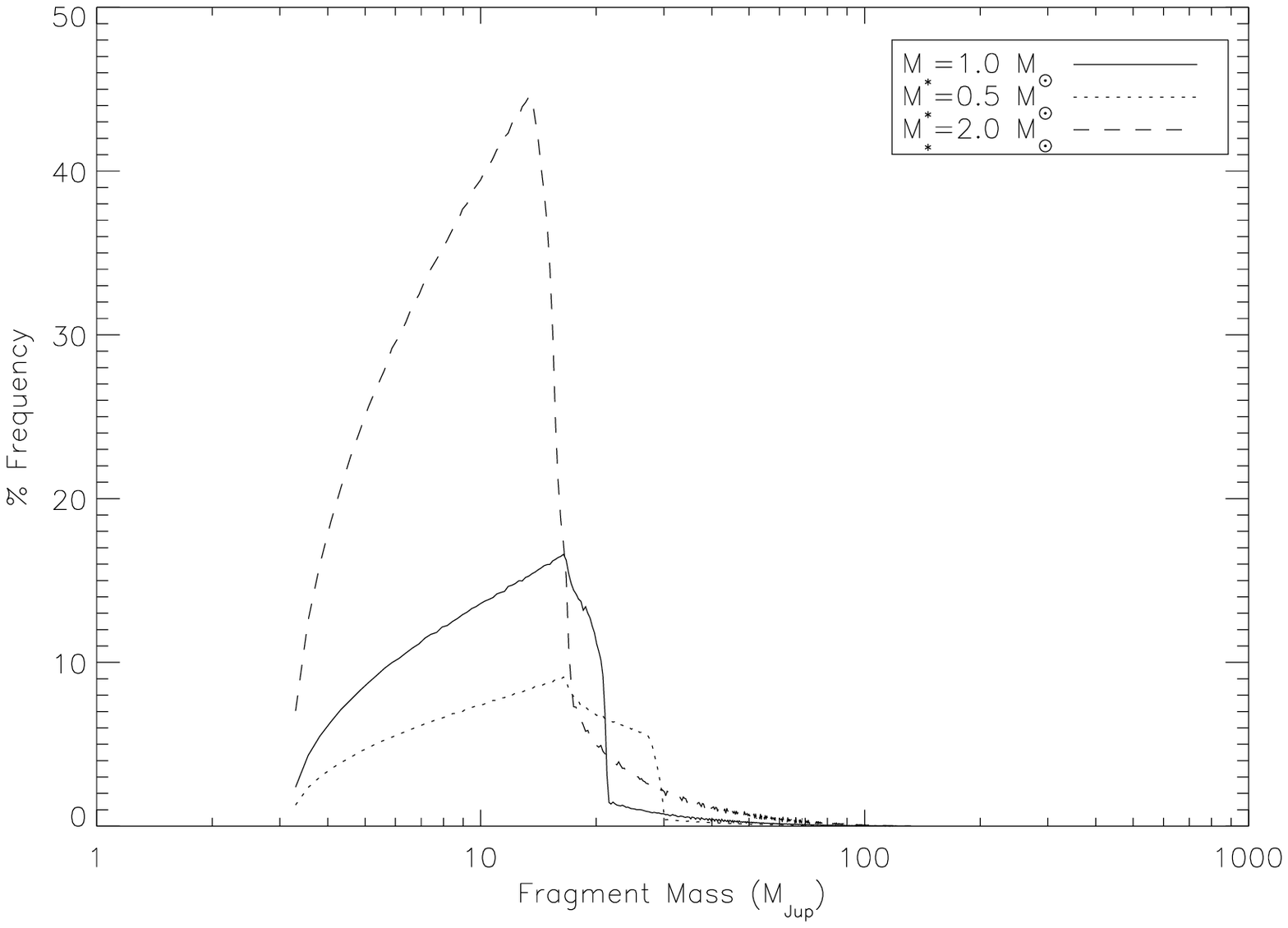} &
\includegraphics[scale = 0.5]{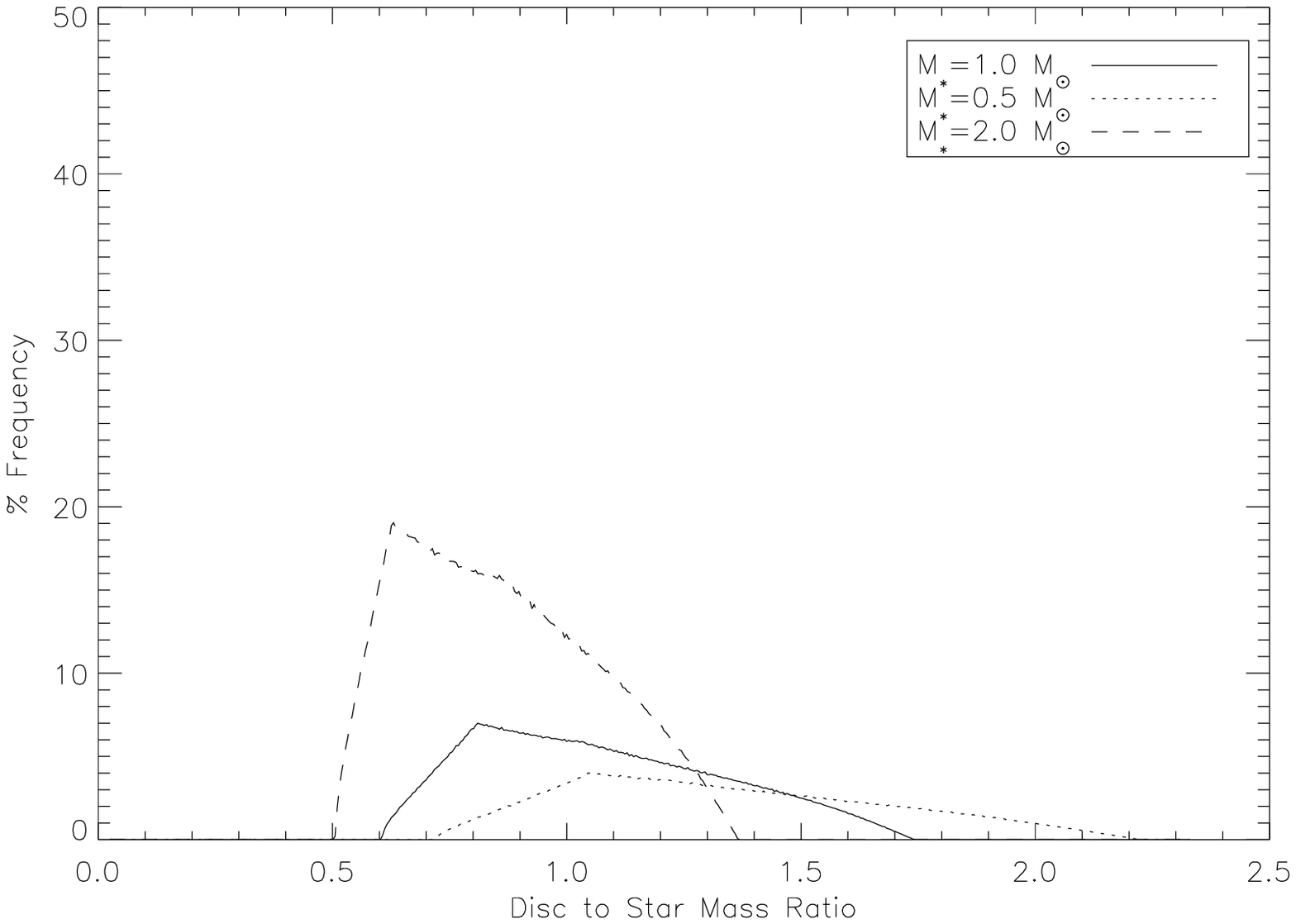} \\

\end{array}$
\caption{The frequency distribution of fragment mass (left) and disc-to-star mass ratio $q$ (right) from disc models which resulted in fragmentation.  Both distributions are binned - the bin widths are $0.2$ $M_{Jup}$ for the fragment mass, and $5 \times 10^{-3}$ for $q$.}\label{fig:bin}
\end{center}
\end{figure*}

What can be said about the expected statistics of these objects? We present binned frequency distributions of the fragment mass from the parameter survey in Figure \ref{fig:bin} (left panel).  While this data will be in no way representative of the true distribution of fragment mass, it indicates that in a uniform disc population, certain fragment masses are considerably more popular than others (given the limits of the parameter space that we have investigated).  In particular, the peak of the distribution occurs between 10 and 20 $M_{\rm Jup}$ for all three stellar masses.  Given that this is the regime in which the giant planet/brown dwarf boundary is located, this has interesting ramifications for planet formation theory, in particular for the canonical disc instability theory \citep{Boss_science}, and the related, more recent tidal downsizing theory \citep{Nayakshin2010,Boley2010b}.  Indeed, if this distribution is convolved with a more realistic distribution of disc parameters, this should provide an initial population of fragment masses.  The onus is then on theory to be able to explain the evolution of this population into the current exoplanet population.  Interestingly, the peak at 10-20 $M_{\rm Jup}$ coincides with the upper limit for efficient core sedimentation derived by \citet{Nayakshin2010a}.  If fragments are more commonly formed within the limits of efficient sedimentation, then it might be reasonable to assume that giant planets are capable of hosting more substantial cores than previously assumed (cf \citealt{D'Angelo2010} and references within).

Lower mass fragments occur more frequently around lower mass stars, which would suggest that if disc instability is to form planets, it will be required to do so at an early stage in protostellar evolution, as suggested by \citet{Greaves2010}.  This is supported by the frequency distribution of $q$ (Figure \ref{fig:bin}, right panel).  Lower mass stars require in general a larger $q$ to form fragments, again suggesting that giant planet formation by disc instability must occur early in the star's evolution.  The minimum $q$ for fragmentation is approximately 0.5, in agreement with the limit set by \citet{Kratter2010a}.  

We must now return to something of a thorny problem in massive, self-gravitating discs - the $\alpha$-parametrisation.  We have demonstrated that fragmentation occurs when $q>0.5$, and we have implicitly assumed a local model of angular momentum transport in our derivation of $\Gamma_J$ and in the one-dimensional disc models.  But such massive self-gravitating discs are not expected to be local \citep{Balbus1999}.  The development of large amplitude, $m=2$ spiral modes in the disc will produce non-local features in the gravitational potential, significantly affecting the stresses developed at any location in the disc.  \citet{Forgan2011} demonstrate that non-local effects become important for isolated discs with $q>0.5$, and also that $Q$ and $T$ become extremely variable with time.  For discs with infall, this limit drops to $q \geq 0.1$ \citep{Harsono2011}.  The implication of this is that we should be cautious in applying the quantitative results from the disc models.  To correct this, we require data on how $\Delta \Sigma/\Sigma$ depends on disc variables when $q>0.5$, in essence to repeat \citet{Cossins2008}'s analysis at higher $q$.  Equally, we must correct a similar deficiency in $\Gamma_J$ by replacing the viscous heating term with something more accurate, perhaps also incorporating the effects of stellar irradiation, an important stabilising process in self-gravitating discs \citep{Rafikov_07,Mejia_4,Stamatellos2008}.  If $Q$ varies significantly, it may be prudent to investigate $\Gamma_J$'s dependence on an extra timescale:

\begin{equation} \beta_Q = \frac{Q}{\dot{Q}} \Omega \end{equation}

\noindent and recalculate $\dot{M}_J$ with the $Q \sim 1$ assumption relaxed.  Indeed, its reliance on such few assumptions lends itself to use in numerical simulations where non-local transport does not dominate, and provides another means of ensuring that simulations satisfy the resolution criteria of \citet{Burkert_Jeans}, which explicitly requires the Jeans mass to be resolved by a sufficient number of SPH particles (or equivalently that the Jeans length be resolved by a sufficient number of grid cells).

While we have focused exclusively on self-gravitating protostellar discs in this paper, the Jeans mass expression permits itself to be taken to any size scale.  For example, this work would apply to the fragmentation of discs around the supermassive black holes (SMBHs) that are expected to reside at the centre of most galaxies.  \citet{Nayakshin2007} use constant $\beta_c$ SPH simulations to investigate star cluster formation (and subsequent star formation) around the Milky Way's own  SMBH, Sgr A*.  They find a different value for $\beta_{\rm crit}$ than is typically found for protostellar discs, and the local initial mass function (IMF) due to disc fragmentation is top-heavy in comparison to the Solar Neighbourhood.  We find that (using their parameters of $M_{\rm bh} = 3.5\times 10^6 M_{\odot}$, and $M_{\rm disc} = 0.01 M_{\rm bh}$), that the Jeans mass is

\begin{equation} M_J = \frac{204}{\left(1 + 1/\sqrt{\beta_c}\right)} M_{\odot}. \label{eq:MJAGN}\end{equation}

\noindent \citet{Nayakshin2007} use $\beta_c =0.2$ as their lowest value, giving a minimum $M_J=63 M_{\odot}$. This is still around an order of magnitude larger than the masses obtained by \citet{Nayakshin2007} and other works in the same field (e.g. \citealt{Bonnell2008}).  It is difficult to compare initial fragment masses with the resulting stellar masses grown through accretion, and it is also true that fragment destruction and mass loss by fragment-fragment collision is more frequent \citep{Levin2007}.  Equally, this larger Jeans mass does suggest that the subsequent IMF will be top-heavy.  An avenue for further work is to apply the expressions for $\Gamma_J$ and $M_J$ in the same fashion as we have done for protostellar discs to yield the initial distribution of fragment masses for a variety of AGN disc parameters, and confirm similar analyses by e.g. \citet{Levin2007} to obtain the expected mass distribution.

We have said much about constant $\beta_c$ simulations in this paper.  While perhaps a little unrealistic, they have been extremely valuable to accretion disc researchers in getting a feel for what disc regimes are likely to produce fragments.  Now that we have introduced a more generalised timescale $\Gamma_J$, can we now consider a set of controlled experimental simulations where $\Gamma_J$ is held constant?  \citet{Mayer2004} perform a set of SPH simulations where $\Gamma_\Sigma$ is controlled by allowing the mass of particles to increase by some prescription.  While they use a more complex cooling function than $\beta_c$, it is possible that the two techniques could be merged to produce a simulation where $\Gamma_J$ is roughly constant at all radii, although incorporating a non-zero $\dot{\beta}_c$ is more challenging.  Developing such constant $\Gamma_J$ simulations to gather empirical data is definitely worth future investigation.

We should finally acknowledge a subtlety regarding the sound speed.  As with the surface density, we should really consider the sound speed inside the spiral wave, which will be enhanced relative to the background due to shocks.  Shocks in self-gravitating protostellar discs typically have Mach numbers $M\sim 1$ \citep{Cossins2008}.  The fractional perturbation amplitude in temperature $\Delta T/T$ in adiabatic shocks at $\gamma = 5/3$ for $M=1$ is approximately 0.3125, implying that our expression for the Jeans Mass will underestimate the true value by roughly 30\%.  Preliminary work on Smoothed Particle Hydrodynamics (SPH) simulations of disc formation from collapsing molecular clouds shows that the fragment masses are systematically under-estimated by a similar factor (Forgan and Rice, in prep.).  In any case, equation (\ref{eq:mjeans_H}) is a \emph{minimum} fragment mass estimate, and should be treated as such, especially when the growth of the fragments must also be considered.  

Indeed, an interesting avenue for further work is to follow the subsequent evolution of these fragments.  Being able to accrete from what is still a substantial pool of gaseous material in the disc will significantly influence their final mass, and the available feedstock for any subsequent planet formation in the inner regions (say by core accretion).  This will surely affect any potential bimodal distribution of planetary parameters \citep{Boley2009}.  As we have said previously, the ability of the fragments to grow as they potentially migrate inwards is of high importance both to canonical disc instability theory and to the more recent tidal downsizing hypothesis \citep{Nayakshin2010a}.  Also, the future interaction of the fragments with the remaining disc will create strong tidal structures, providing pressure maxima for dust grains to occupy and potentially enhance their growth rate \citep{Rice2004,Clarke2009}.  The expressions for $M_J$ and $\Gamma_J$ will be of use for further work in this two-phase regime.

\section{Conclusions }\label{sec:conclusions}

\noindent We have derived the local Jeans mass inside a spiral perturbation, and shown that it can be calculable using azimuthally averaged disc variables.  This allows the expected mass of objects formed from disc fragmentation to be calculated simply, both in theoretical models and in numerical simulations, across a variety of size scales.  We have taken this expression and derived a Jeans timescale $\Gamma_J$, which describes the length of time required for the Jeans Mass to increase or decrease by its current value (normalised by the local angular velocity, in much the same way as the cooling time is often normalised).  We show that the resulting expression for $\Gamma_J$ reduces to the standard cooling time criterion as was previously defined \citep{Gammie,Rice_et_al_05}, and folds in related results on the disc's thermal history \citep{Clarke2007} and envelope accretion \citep{Kratter2010a}.  We confirm that subjecting the disc to extra shock heating promotes fragmentation, and that rapidly accretion encourages discs to fragment, provided that local angular momentum transport and cooling is efficient.

We have investigated the functional form of the Jeans Mass by applying it to simple one-dimensional steady-state self-gravitating disc models.  We find that the minimum fragment mass is around 3 Jupiter Masses, in agreement to previous work on the subject \citep{Rafikov_05,Kratter2009,Nero2009}.  We also find that in a uniform survey of disc parameters, the most frequent fragment masses lie between 10 - 20 $M_{\rm Jup}$.  These results are relatively insensitive to stellar mass, although lower mass stars require a higher disc-to-star mass ratio to produce fragments than their higher-mass counterparts.  Conversely, the mass of fragments found around lower mass stars are higher than that for higher mass stars (for the same outer radius and steady-state accretion rate).  We do however note that the derivation of the Jeans mass relies on assuming local angular momentum transport, which is often not the case for discs of high disc-to-star mass ratio.  The errors introduced by this assumption require further empirical data about non-local angular momentum transport to quantify accurately. 

The formalisms introduced in this paper are amenable to a variety of studies in self-gravitating discs of all scales, allowing the intial fragment mass distribution to be identified given the known disc population, and to provide more general criteria for disc fragmentation itself.

\section*{Acknowledgments}

\noindent DF and KR gratefully acknowledge support from STFC grant ST/H002380/1.

\bibliographystyle{mn2e} % (must include a bibliography style)
\bibliography{jeans_selfgravdisc}

\appendix

\label{lastpage}

\end{document}